\documentclass[10pt,conference]{IEEEtran}
\IEEEoverridecommandlockouts
\usepackage{cite}
\usepackage{amsmath,amssymb,amsfonts}
\usepackage{algorithmic}
\usepackage{graphicx}
\usepackage{textcomp}
\usepackage{xcolor}
\usepackage{comment}
\usepackage{booktabs}
\usepackage{tcolorbox}
\usepackage{array,multirow,tabularx,ragged2e}
\PassOptionsToPackage{hyphens}{url}\usepackage{hyperref} 
\usepackage{pifont}
\usepackage{array}
\usepackage{multirow}
\usepackage{subfig}
\usepackage{xspace}
\usepackage{listings}

\newcommand{\symmetric}{Symmetric}
\newcommand{\asymmetric}{Asymmetric}
\newcommand{\hash}{Hash}
\newcommand{\random}{Random}

\newcommand\CCFinderGroups{2,657}
\newcommand\RelevantGroups{953}
\newcommand\IrrelevantGroups{1,704}
\newcommand\SecureGroups{587}
\newcommand\InsecureGroups{326}
\newcommand\MixedGroups{40}
\newcommand\RelevantInstances{3,121}
\newcommand\SecureInstances{1,802}
\newcommand\InsecureInstances{1,319}
\newcommand\IrrelevantInstances{5,569}
\newcommand\AllInstances{8,690}
\newcommand\TotalSnippet{25,855}
\newcommand\SecureAnswers{785}
\newcommand\InsecureAnswers{644}
\newcommand\AllAnswers{3,562}
\newcommand\TotalAnswers{23,224}
\newcommand\ExcludedAnswers{19,662}
\newcommand\RelevantAnswers{1,429}
\newcommand\IrrelevantAnswers{2,133}
\newcommand\RelevantAnswersSSL{505}
\newcommand\SecureAnswersSSL{150}
\newcommand\InsecureAnswersSSL{355}

\newcommand\SumCategoryAnswers{1,506}
\newcommand\AverViewsSecureAnswers{18,713}
\newcommand\AverViewsInsecureAnswers{36,508}
\newcommand\AverScoreSecureAnswers{5}
\newcommand\AverScoreInsecureAnswers{14}
\newcommand\AverCommentsSecureAnswers{2}
\newcommand\AverCommentsInsecureAnswers{3}
\newcommand\AverReputationSecureAnswers{18,654}
\newcommand\AverReputationInsecureAnswers{14,678}
\newcommand\AverFavoritesSecureAnswers{8}
\newcommand\AverFavoritesInsecureAnswers{15}

\newcommand\HighestReputation{990,402}
\newcommand\LowestReputation{1}
\newcommand\RelevantAnswersByTrustedUsers{208}

\newcommand\InsecureAnswersByTrustedUsers{71}
\newcommand\AcceptedSecureAnswers{297}
\newcommand\AcceptedInsecureAnswers{239}
\newcommand\AcceptedRelevantAnswers{536}
\newcommand\HighReputationInsecureAnswers{26}
\newcommand\HighReputationRelevantAnswers{72}

\newcommand\AnswerReusageUsers{109}
\newcommand\SecureReusageUsers{66}
\newcommand\InsecureReusageUsers{49}

\newcommand\SecureGroupsOfReusedAnswers{111}
\newcommand\InsecureGroupsOfReusedAnswers{90}
\newcommand\MixedGroupsOfReusedAnswers{6}
\newcommand\RelevantGroupsOfReusedAnswers{207}

\newcommand\SecureGroupMeanSize{3.0}
\newcommand\InsecureGroupMeanSize{3.8}
\newcommand\GroupSizePValue{1.3e-4}
\newcommand\GroupSizeCliff{0.13}

\newcommand\ReusedPairsForDiffQ{46}
\newcommand\SecureSamplePairs{56}
\newcommand\MixedSamplePairs{2}
\newcommand\InsecureSamplePairs{45}

\newcommand{\so}{SO\xspace}
\newcommand{\SO}{SO\xspace}

\newcommand{\cmark}{\ding{51}}%

\newcommand{\codefont}[1]{\footnotesize{\texttt{#1}}\normalsize}

\newcolumntype{L}[1]{>{\raggedright\let\newline\\\arraybackslash\hspace{0pt}}m{#1}}
\newcolumntype{C}[1]{>{\centering\let\newline\\\arraybackslash\hspace{0pt}}m{#1}}
\newcolumntype{R}[1]{>{\raggedleft\let\newline\\\arraybackslash\hspace{0pt}}m{#1}}

\def\BibTeX{{\rm B\kern-.05em{\sc i\kern-.025em b}\kern-.08em
    T\kern-.1667em\lower.7ex\hbox{E}\kern-.125emX}}
\begin{document}

\title{How Reliable is the Crowdsourced Knowledge of Security Implementation?
\footnotesize
\thanks{This work was supported by ONR Grant N00014-17-1-2498.}
}

\author{\IEEEauthorblockN{Mengsu Chen$^{\ast}$ \hspace*{3em} Felix Fischer$^{\dag}$ \hspace*{3em} Na Meng$^{\ast}$ \hspace*{3em} Xiaoyin Wang$^{\ddag}$ \hspace*{3em} Jens Grossklags$^{\dag}$ \hspace*{3em}}
\IEEEauthorblockA{
Virginia Tech$^{\ast}$ \hspace*{3em} Technical University of Munich$^{\dag}$ \hspace*{3em}
University of Texas at San Antonio$^{\ddag}$\\ 
mschen@vt.edu, flx.fischer@tum.de, nm8247@vt.edu, xiaoyin.wang@utsa.edu, jens.grossklags@tum.de}
}

\maketitle

\begin{abstract}

Stack Overflow (\SO) 
is the most popular online 
Q\&A site for developers to share their expertise in solving programming issues. 
Given multiple answers to certain questions, developers 
may take the accepted answer, the answer from a person with high reputation, or the one frequently suggested. 
However, researchers recently observed that \SO contains exploitable security vulnerabilities in the suggested code of popular answers, which found their way into security-sensitive high-profile applications that millions of users install every day.
This observation inspires us to explore the following questions: How much can we trust the security implementation suggestions on \SO? If suggested answers are vulnerable, can developers rely on the community's dynamics to infer the vulnerability and identify a secure counterpart?


To answer these highly important questions, we conducted a comprehensive study on security-related \SO posts by contrasting secure and insecure advice with the community-given content evaluation. Thereby, we investigated whether \so{}'s gamification approach on incentivizing users is effective in improving security properties of distributed code examples. Moreover, we traced the distribution of duplicated samples over given answers to test whether the community behavior facilitates or prevents propagation of secure and insecure code suggestions within \so{}.

We compiled \RelevantGroups\ different groups of similar security-related code examples
and labeled their security, identifying \SecureAnswers\ secure answer posts and \InsecureAnswers\ insecure answer posts.  
Compared with secure suggestions, insecure ones had higher view counts (\AverViewsInsecureAnswers\ vs.~\AverViewsSecureAnswers),
received a higher score (\AverScoreInsecureAnswers\ vs.~\AverScoreSecureAnswers),
and had significantly more duplicates (\InsecureGroupMeanSize\ vs.~\SecureGroupMeanSize) on average. 34\% of the posts provided by highly reputable so-called \textit{trusted users} were insecure.  

Our findings show that based on the distribution of secure and insecure code on \so{}, users being laymen in security rely on additional advice and guidance.
However, the community-given feedback does not allow  differentiating secure from insecure choices. The reputation mechanism fails in indicating trustworthy users with respect to security questions, ultimately leaving other users wandering around alone in a software security minefield.

\end{abstract}


\section{Introduction}
\label{sec:introduction}
Since its launch in 2008, Stack Overflow (\SO) has served as the infrastructure for developers to discuss programming-related questions online, and provided the community with crowdsourced knowledge~\cite{so_crowdsourced,so_crowdsourced_2010}. Prior work shows that \SO is one of the most important information resources that developers rely on~\cite{mamykina_design_2011,Acar:2016}. Meanwhile, researchers also revealed that some highly upvoted,
or even accepted answers on \SO contained insecure code
~\cite{fischer:2017,Meng2018:icse}. 
More alarmingly, Fischer \emph{et al.}~found that insecure code snippets from \SO were copied and pasted into 196,403 Android applications available on Google Play~\cite{fischer:2017}. 
Several high-profile applications containing particular instances of these insecure snippets were successfully attacked, and user credentials, credit card numbers and other private data were stolen as a result~\cite{Fahl:2012}.

These observations made us curious about SO's reliability regarding suggestions for security implementations. 
Taking a pessimistic view, such insecure suggestions can be expected to be prevalent on the Q\&A site, and consistent corrective feedback by the programming community may be amiss. Consequently, novice developers may learn about incorrect crowdsourced knowledge from such Q\&A sites, propagate the misleading information to their software products or other developers, and eventually make our software systems vulnerable to known security attacks.

Therefore, within this paper, we conducted \emph{a comprehensive in-depth investigation of the popularity of both secure and insecure coding suggestions on \SO, and the community activities around them}. To ensure a fair comparison between secure and insecure suggestions, we focused on the discussion threads related to Java security. 
We used Baker~\cite{Subramanian:2014} to mine for answer posts that contained any code using security libraries, and extracted 25,855 such code snippets.
We reused the security domain expertise summarized by prior work~\cite{fischer:2017} to manually label whether a given code snippet is secure or not. 
However, different from prior work~\cite{fischer:2017} that studied the application of insecure \SO answers to production code, 
our work focuses on the SO suggestions themselves. 
More specifically, we studied coding suggestions' popularity, social dynamics, and duplication. We also inquired 
how developers 
may be misled by insecure answers on \SO. 

To identify prevalent topics on \so{}, we used CCFinder~\cite{Kamiya2002} to detect code clones (\emph{i.e.}, duplicated code) in the data extracted by Baker. These clones are clustered within clone groups. \RelevantGroups\ clone groups were observed to use security library APIs and implement functionalities like SSL/TLS, symmetric and asymmetric encryption, \emph{etc.} 
Moreover, we found that code examples within clone groups are more likely to be viewed than non-duplicated code snippets on \so{}. This further motivates our sampling method as we can expect clones to have a higher impact on users and production code. 
Among the \RelevantGroups\ clone groups, 
there were \SecureGroups\ groups of duplicated secure code,
\InsecureGroups\ groups of similar insecure code, and \MixedGroups\ groups with a mixture of secure and insecure code snippets. 
These clone groups covered \SecureInstances\ secure code snippets and \InsecureInstances\ insecure ones. 
By mapping cloned code to their container posts, 
we contrasted insecure suggestions with secure ones in terms of their popularity, users' feedback, degree of duplication, and causes for duplication. 

We explored the following Research Questions (RQs):

\begin{itemize}
\setlength\itemsep{0.5em}
\item \textbf{RQ1:} \emph{How prevalent are insecure coding suggestions on \SO?} 
Prior work witnessed the existence of vulnerable code on \SO, and indicates that such code can mislead developers and compromise the quality of their software products~\cite{Acar:2016,fischer:2017,Meng2018:icse}. To understand the number and growth of secure and insecure options that developers have to choose from, we (1) compared the occurrence counts of insecure and secure answers, and (2) observed the distributions of both kinds of answers across a 10-year time frame (2008-2017).

\item \textbf{RQ2:} \emph{Do the community dynamics or \SO's reputation mechanism help developers choose secure answers over insecure ones?}
Reputation mechanisms and voting were introduced to crowdsourcing platforms to (1) incentivize contributors to provide high-quality solutions, and (2) facilitate question askers to identify responders with high expertise~\cite{Jurczyk:2007,Xie2015,Katmada2016}.
We conducted statistical testing to compare secure and insecure answers in terms of votes, answerers' reputations, \emph{etc.} 


\item \textbf{RQ3:} \emph{Do secure coding suggestions have more  duplicates than insecure ones?}
When certain answers are repetitively suggested, it is 
likely that developers will encounter such answers more often. Moreover, if these answers are provided by different users, the phenomenon might facilitate users' trust in the answers' correctness.
Therefore, we compared the degree of repetitiveness for insecure and secure answers.

\item \textbf{RQ4:} \emph{Why did users suggest duplicated secure or insecure answers on \SO?}
We were curious about why certain code was repetitively suggested, and we explored this facet of community behavior by examining the duplicated answers posted by the same or different users.

\end{itemize}

In our study, we made four major observations:
\begin{enumerate}
\setlength\itemsep{0.5em}
\item \emph{As with secure answers, insecure answers are prevalent on \SO across the entire studied time frame.} 
The inspected \RelevantInstances\ snippets from different clone groups correspond to \SecureAnswers\ secure posts and \InsecureAnswers\ insecure ones.  
Among the \RelevantAnswersSSL\ SSL/TLS-related posts, \InsecureAnswersSSL\ posts (70\%) suggest insecure solutions, which makes SSL/TLS-related answers the most unreliable ones on \so{}.
At least 41\% of the security-related answers posted every year are insecure, which shows that security knowledge on \so{} in general is not significantly improving over time. 

\item \emph{The community dynamics and SO's reputation mechanisms are not reliable indicators for secure and insecure answers.} 
Compared with secure posts, insecure ones obtained higher \emph{scores}, more \emph{comments}, more \emph{favorites}, and more \emph{views}. 
Although the providers of secure answers received significantly higher \emph{reputation} scores, the effect size is negligible ($<$0.147). 
\AcceptedInsecureAnswers\ of the \AcceptedRelevantAnswers\ examined \emph{accepted answers} (45\%) are insecure. 
\HighReputationInsecureAnswers\ out of the \HighReputationRelevantAnswers\ posts (36\%) suggested by ``trusted users'' (with $\ge$20K reputation scores~\cite{so-priviledges}) are insecure.
These observations imply that reputation and voting on \so{} are not reliable to help users distinguish between secure and insecure answers. 

\item 
\emph{The degree of duplication among insecure answers is significantly higher than that of secure ones.} 
On average, 
there are more clones in an insecure group than a secure one (\InsecureGroupMeanSize\ vs.~\SecureGroupMeanSize). It means that users may have to deal with a large supply of insecure examples for certain questions, before obtaining secure solutions. 

\item \emph{Users seem to post duplicated answers, while ignoring security as a key property.} 
Duplicated answers were provided due to duplicated questions or users' intent to answer more questions by reusing code examples.
This behavior is incentivized by the reputation system on \so{}. The more answers are posted by a user and up-ranked by the community, the higher reputation the user gains.


\end{enumerate}
Our source code and data set are available at \url{https://github.com/mileschen360/Higgs}. 

\section{Background}
\label{sec:background}

To facilitate the discussion of \SO community activities around security implementations, we will first introduce \SO's crowdsourcing model, and then summarize the domain knowledge used to label secure and insecure code snippets. 

\begin{figure}
\includegraphics[width=9cm]{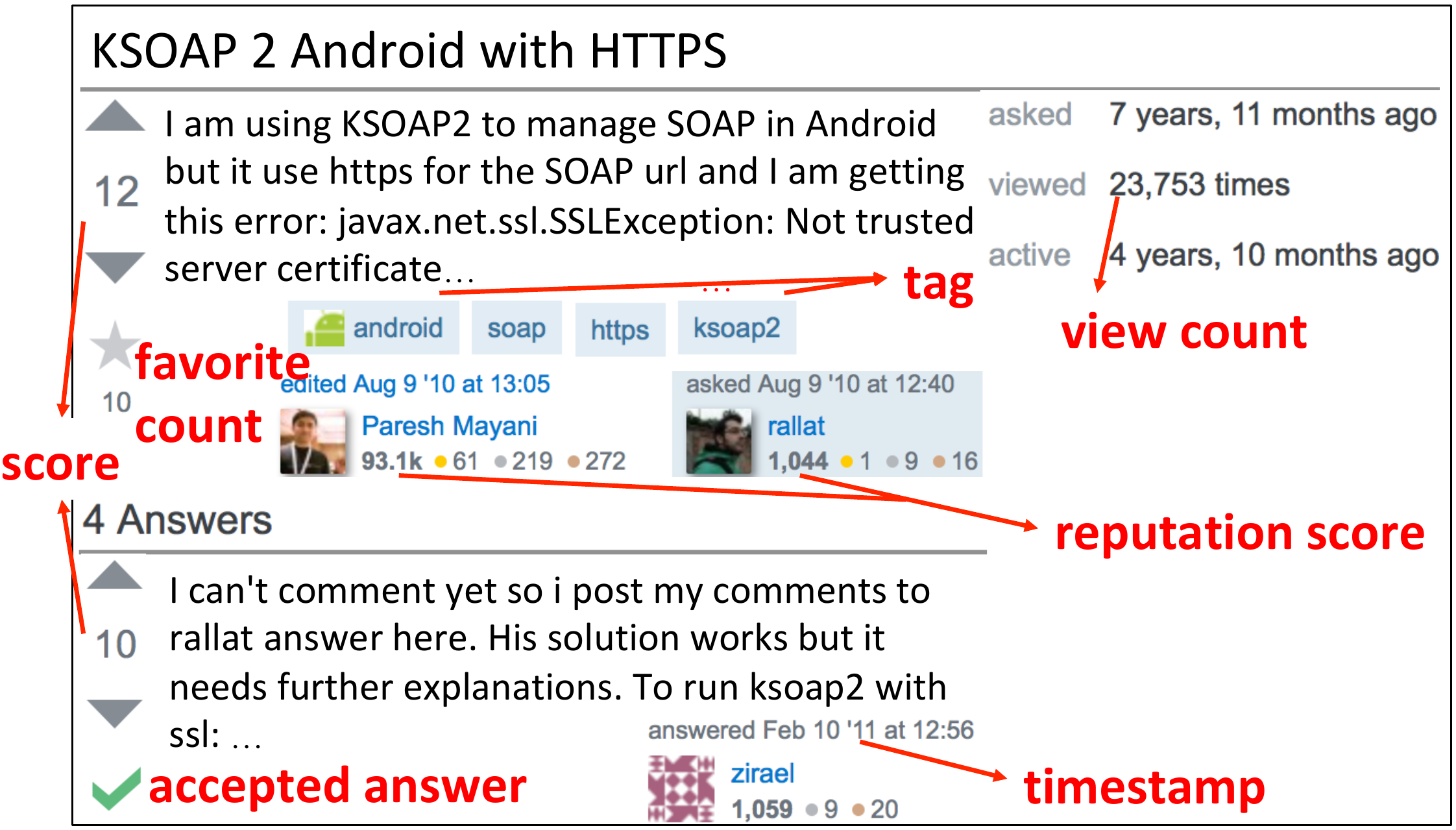}
\caption{A typical \SO discussion thread contains one question post and one or multiple answer posts~\cite{exemplarso}}
\label{fig:so}
\vspace{-1.5em}
\end{figure}
\subsection{Stack Overflow as a Crowdsourcing Platform}
Some observers believe that the success of \SO lies in its crowdsourcing model and the reputation system~\cite{so_crowd,so_crowdsourced_2010}. 
The four most common forms of participation are i) question asking, ii) question answering, iii) commenting, and iv) voting/scoring~\cite{so_crowd}. 
Figure~\ref{fig:so} presents an exemplar \SO discussion thread, which contains one question and one or many answers.  
When multiple answers are available, the asker decides which answer to \textbf{accept}, and marks it with ``\textcolor{green}{\cmark}''. 

After a user posts a question, an answer, or a comment, other users can vote for or against the post. 
Users gain \textbf{reputation} for each up-vote their posts receive. 
For instance, answers earn their authors 10 points per up-vote, questions earn 5, and comments earn 2~\cite{what-reputation}. 
All users initially have only one reputation point. 
As users gain more reputation, they are granted more administrative privileges to help maintain \SO posts~\cite{so-priviledges}. For instance, a user with 15 points can vote up posts. A user with 125 points can vote down posts. Users with at least 20K points are considered ``\textbf{trusted users}'', and can edit or delete other people's posts. 

The \textbf{score} of a question or answer post is decided by the up-votes and down-votes the post received. 
Users can \textbf{favorite} a question post if they want to bookmark the question and keep track of any update on the discussion thread. 
Each question contains one or many \textbf{tags}, which are words or phrases to describe topics of the question.
Each post has a \textbf{timestamp} to show
when it was created. 
Each discussion thread has a \textbf{view count} to indicate how many times the thread has been viewed.

\subsection{Categorization of Java Security Implementations}
Based on state-of-the-art security knowledge, researchers defined five categories of security issues relevant to library misuses~\cite{fischer:2017}.
Table~\ref{tab:rules} shows their criteria, which we use in this project to decide whether a code snippet is insecure or not. 

\begin{table}
\caption{Criteria used to decide code's security property}
\label{tab:rules}
\vspace{-0.5em}
\scriptsize
\begin{tabular}{C{1.5cm}|C{2.3cm}|C{3.7cm}}
\toprule
\textbf{Category} & \textbf{Parameter} & \textbf{Insecure} \\ 
\toprule
\multirow{5}{*}{\textbf{SSL/TLS}} & HostnameVerifier & allow all hosts
\\ \cline{2-3}
& Trust Manager & trust all 
\\ \cline{2-3}
& Version & $<$TLSv1.1 
\\ \cline{2-3}
& {Cipher Suite} & RC4, 3DES, AES-CBC
MD5, MD2\\ \cline{2-3}
& {OnReceivedSSLError} & proceed
\\ \toprule
& {Cipher/Mode} & RC2, RC4, DES, 3DES, AES/ECB, Blowfish\\ \cline{2-3}
\textbf{\symmetric} & {Key} & static, bad derivation\\ \cline{2-3}
& Initialization Vector (IV) & zeroed, static, bad derivation\\ \cline{2-3} 
& Password Based Encryption (PBE) & $<$1k iterations, $<$64-bit salt, static salt  
\\ \toprule
\textbf{Asymmetric} 
& Key &RSA $<$ 2,048 bit, ECC $<$ 224 bit 
 \\ \toprule
& PBKDF & $<$SHA224, MD2, MD5 \\ \cline{2-3}
\textbf{\hash}& Digital Signature & SHA1, MD2, MD5\\ \cline{2-3} 
& Credentials & SHA1, MD2, MD5\\ \toprule
& Type & Random \\ \cline{2-3}
\textbf{\random }& Seeding & setSeed$\rightarrow$nextBytes, setSeed with static values 
\\ 
\bottomrule
\end{tabular}
\vspace{-1.5em}
\end{table}

\paragraph*{\textbf{SSL/TLS}}
There are five key points concerning how to \emph{securely} establish SSL/TLS connections. First, developers should use an implementation of the \codefont{HostnameVerifier} interface
to verify servers' hostnames instead of allowing all hosts~\cite{Fahl:2012}. 
Second, when implementing a custom \codefont{TrustManager}, developers should validate certificates instead of blindly trusting all certificates. 
Third, when ``TLS'' is passed as a parameter to  \codefont{SSLContext.getInstance(...);}, developers should explicitly specify the version number to be at least 1.1, because TLS' lower versions are insecure~\cite{Sheffer2015}.
Fourth, the usage of insecure cipher suites should be avoided. Fifth, 
when overriding \codefont{onReceivedSslError()}, developers should handle instead of skipping any certificate validation error. 
Listing~\ref{lst:ssl} shows a vulnerable snippet that allows all hosts, trusts all certificates, and uses TLS v1.0.

\lstset{
basicstyle=\scriptsize,
breaklines=true,
escapeinside={(*}{*)},
frame = tb}
\begin{lstlisting}[caption=An example to demonstrate 
three scenarios of insecurely using SSL/TLS APIs~\cite{trust-all-allow-all}, label={lst:ssl}]
(*\bfseries // Create a trust manager that does not validate certificate chains (trust all)*)
private TrustManager[] trustAllCerts = new TrustManager[] {
  new X509TrustManager() {
    public java.security.cert.X509Certificate[] getAcceptedIssuers() {return null;}
    public void checkClientTrusted(...) {}
    public void checkServerTrusted(...) {}    }};    
public ServiceConnectionSE(String url) throws IOException {
  try {
    (*\bfseries // Use the default TLSv1.0 protocol *)
    SSLContext sc = SSLContext.getInstance("TLS");
    (*\bfseries // Install the trust-all trust manager*)
    sc.init(null, trustAllCerts, new java.security.SecureRandom()); ... } ...
  connection = (HttpsURLConnection) new URL(url).openConnection();
  (*\bfseries // Use AllowAllHostnameVerifier that allows all hosts *)
  ((HttpsURLConnection) connection).setHostnameVerifier(new AllowAllHostnameVerifier());    }  
\end{lstlisting}

\paragraph*{\textbf{\symmetric}} 
There are ciphers and modes of operations known to be insecure. Cryptographic keys and initialization vectors (IV) are insecure if they are statically assigned, zeroed, or directly derived from text. Password Based Encryption (PBE) is insecure if the iteration number is less than 1,000, the salt's size is smaller than 64 bits, or a static salt is in use. Listing~\ref{lst:symmetric} presents a vulnerable code example that insecurely declares a cipher, a key, and an IV. 

\begin{lstlisting}[caption=An example to present 
several 
insecure usage scenarios of symmetric cryptography~\cite{keyiv}, label=lst:symmetric]
(*\bfseries // Declare a key parameter with a static value*)
private static byte[] key = "12345678".getBytes();
(*\bfseries // Declare an IV parameter with a static value*)
private static byte[] iv = "12345678".getBytes();
public static String encrypt(String in) {
  String cypert = in;
  try {
    IvParameterSpec ivSpec = new IvParameterSpec(iv);
    (*\bfseries // Create a secret key with the DES cipher*)
    SecretKeySpec k = new SecretKeySpec(key, "DES");
    (*\bfseries // Declare a DES cipher*)
    Cipher c = Cipher.getInstance("DES/CBC/PKCS7Padding");
    c.init(Cipher.ENCRYPT_MODE, k, ivSpec);
    ... }    }
\end{lstlisting}

\paragraph*{\textbf{\asymmetric}}
Suppose that a code snippet uses either RSA or ECC APIs to generate keys. 
When the specified key lengths for RSA and ECC are separately shorter than 2,048 bits and 224 bits, we consider the API usage to be insecure. Listing~\ref{lst:asymmetric} shows a vulnerable code example. 

\begin{lstlisting}[caption=An example that insecurely uses RSA by specifying the key size to be 1024~\cite{asymmetric}, label=lst:asymmetric]
KeyPairGenerator kpg = KeyPairGenerator.getInstance("RSA");
kpg.initialize(1024);
KeyPair kp = kpg.generateKeyPair();
RSAPublicKey pub = (RSAPublicKey) kp.getPublic();
RSAPrivateKey priv = (RSAPrivateKey) kp.getPrivate();
\end{lstlisting}

\paragraph*{\textbf{\hash}}
In the context of password-based key derivation, digital signatures, and authentication/authorization, developers may explicitly invoke broken hash functions. Listing~\ref{lst:hash} shows an example using MD5.

\begin{lstlisting}[caption=Insecurely creating a message digest with MD5~\cite{md5}, label=lst:hash]
final MessageDigest md = MessageDigest.getInstance("md5");
(*\bfseries // It is also insecure to hardcode the plaintext password*)
final byte[] digestOfPassword = md.digest("HG58YZ3CR9".getBytes("utf-8"));
\end{lstlisting}

\paragraph*{\textbf{\random}}
To make the generated random numbers unpredictable and secure, developers should use \codefont{SecureRandom} instead of \codefont{Random}. When using \codefont{SecureRandom}, developers can either (1) call \codefont{nextBytes()} only, or (2) call \codefont{nextBytes()} first and \codefont{setSeed()} next. Developers should not call \codefont{setSeed()} with static values. Listing~\ref{lst:sr} presents an example using \codefont{SecureRandom} insecurely. 

\begin{lstlisting}[caption=Using \codefont{SecureRandom} with a static seed~\cite{sr}, label=lst:sr]
byte[] keyStart = "encryption key".getBytes();
SecureRandom sr = SecureRandom.getInstance("SHA1PRNG");
sr.setSeed(keyStart);
\end{lstlisting} 

\vspace{-0.5em}
\section{Methodology}
\label{sec:method}

To collect secure and insecure answer posts, 
we first extracted code snippets from \SO that used any security API (Section~\ref{sec:extract}).
Next, we sampled the extracted code corpus by detecting duplicated code (Section~\ref{sec:clone}).
Finally, we manually labeled sampled code as secure, insecure, or irrelevant, and mapped the code to related posts (Section~\ref{sec:label}). 
Additionally, 
we compared the view counts of the sampled posts vs.~unselected posts to check samples' prevalence
(Section~\ref{sec:cdf}).


\subsection{Code Extraction}
\label{sec:extract}
To identify coding suggestions, this step extracts security-related answer posts by analyzing (1) tags of question posts, and (2) the code snippets' API usage of answer posts.
After downloading the Stack Overflow data as XML files~\cite{datadump},  
we used a tool stackexchange-dump-to-postgres~\cite{xmltotable} to convert the XML files to Postgres database tables. 
Each row in the database table ``Posts'' corresponds to one post. 
A post's body text may use the HTML tag pair 
\codefont{<code>} and \codefont{</code>} to enclose source code, so we leveraged this tag pair to extract code. 
Since there were over 40 million posts under processing, and one post could contain multiple code snippets, 
\emph{it is very challenging to efficiently identify security implementations from a huge amount of noisy code data.} 
Thus, we built two heuristic filters to quickly skip irrelevant posts and snippets. 

\begin{table}[!h]
\caption{Tags used to locate relevant \SO discussion threads}
\label{tab:tags}
\vspace{-0.5em}
\begin{tabular}{C{2cm}| p{6cm}}
\toprule
\textbf{Category} & \textbf{Tags}\\
\toprule
\textbf{Java platforms}
& android, applet, eclipse, java, java1.4, java-7, java-ee, javamail, jax-ws, jdbc, jndi, jni, ...
\\ \hline 
\textbf{Third-party tools/libraries} & axis2, bouncycastle, gssapi, httpclient, java-metro-framework, openssh, openssl, spring-security, ...\\ \hline
\textbf{Security} & aes, authentication, certificate, cryptography, des, encoding, jce, jks, jsse, key, random, rsa, security, sha, sha512, single-sign-on, ssl, tls, X509certificate, ...\\ 
\bottomrule
\end{tabular}
\vspace{-1em}
\end{table}

\subsubsection{Filtering by question tags}
As tags are defined by askers to describe the topics of questions, we relied on tags to skip obviously irrelevant posts. To identify as many security coding suggestions as possible, 
we inspected the 64 cryptography-related posts mentioned in prior work~\cite{Meng2018:icse}, and identified 93 tags. 
If a question post contains any of these tags, we extracted code snippets from the corresponding answer posts.
As shown in Table~\ref{tab:tags}, these tags are either related to Java platforms, third-party security libraries or tools, or security concepts. 

\subsubsection{Filtering by security API usage}
Similar to prior work~\cite{fischer:2017}, we used Baker~\cite{Subramanian:2014} to decide whether a snippet calls any security API. 
This paper focuses on the following APIs: 
\begin{itemize}
\item Java platform security: org.jetf.jgss.*, android.security.*, com.sun.security.*, java.security.*, javax.crypto.*, javax.net.ssl.*, javax.security.*, javax.xml.crypto.*; 
\item Third-party security libraries: 
BouncyCastle~\cite{bc},
GNU Crypto~\cite{gnu-crypto},
jasypt~\cite{jasypt}, 
keyczar~\cite{keyczar}, 
scribejava~\cite{scribejava},
SpongyCastle~\cite{sc}.
\end{itemize}
After taking in a set of libraries and a code snippet, Baker (1) extracts all APIs of types, methods, and fields from the libraries, (2) extracts names of types, methods, and fields, used in the code snippet, and (3) iteratively deduces identifier mappings between the extracted information.
Intuitively, when multiple type APIs (\emph{e.g.}, \codefont{a.b.C} and \codefont{d.e.C}) can match a used type name \codefont{C}, Baker compares the invoked methods on \codefont{C} against the method APIs declared by each candidate type, and chooses the candidate that contains more matching methods. 

We included a code snippet if Baker finds at least one API (class or method) with an exact match. However, Baker's result set is not fully accurate and requires a number of post-processing steps to reduce false positives. These include a blacklist filter for standard Java types (e.g., String) and very popular methods (e.g., get()). Baker's oracle contains only the given security APIs, which helped reduce false positives when detecting secure code but did not help reduce false negatives.  

\subsection{Clone Detection}
\label{sec:clone}
With the filters described above, we identified \TotalSnippet\ code snippets (from \TotalAnswers\ posts) that are probably security-related. 
Since it is almost impossible to manually check all these snippets to identify secure and insecure code, we decided to (1) sample representative extracted code via clone detection, and then (2) manually label the samples. 
In addition to sampling, clone detection facilitates our research in two further ways. 
First, by identifying duplicated code with CCFinder~\cite{Kamiya2002}, we could explore the degree of duplication among secure and insecure code. 
Second, through clustering code based on their similarity, we could efficiently read similar code fragments, and determine their security property in a consistent way. 
With the default parameter setting in CCFinder, we identified \CCFinderGroups\ clone groups that contained \AllInstances\ code snippets, with each group having at least two snippets. 

\subsection{Code Labeling}
\label{sec:label}
We manually examined each of those \AllInstances\ snippets and labeled code based on the criteria mentioned in Section~\ref{sec:background}.
If a snippet meets any criteria of insecure code shown in Table~\ref{tab:rules}, it is labeled as ``insecure''. If the snippet uses any security API but does not meet any criteria, it is labeled as ``secure''; otherwise, it is ``irrelevant''. 
Depending on the APIs involved, we also decided to which security category a relevant post belongs. 
When unsure about certain posts, we had discussions to 
achieve consensus. 
Finally, we randomly explored a subset of the labeled data to double check the correctness. 

\begin{table}
\caption{Code labeling results for \CCFinderGroups\ clone groups}
\scriptsize
\centering
\label{tab:label}
\begin{tabular}{c| rrrr|r}
\toprule
 & \textbf{Secure} & \textbf{Insecure} & \textbf{Mixed} & \textbf{Irrelevant} & \textbf{Total}\\
\toprule
\textbf{\# of clone groups} &\SecureGroups\ & \InsecureGroups\ & \MixedGroups\ & \IrrelevantGroups\ & \CCFinderGroups\ \\ 
\textbf{\# of snippets} &\SecureInstances\ & \InsecureInstances\ & 0 &  \IrrelevantInstances\ & \AllInstances\ \\
\textbf{\# of answer posts} &\SecureAnswers\ & \InsecureAnswers\ & 0 & \IrrelevantAnswers\ & \AllAnswers\ \\ 
\bottomrule
\end{tabular}
\vspace{-0.5em}
\end{table}

Table~\ref{tab:label} presents our labeling results for the inspected \CCFinderGroups\ clone groups. After checking individual code snippets, we identified \SecureGroups\ secure groups, \InsecureGroups\ insecure groups, \MixedGroups\ mixed groups, and \IrrelevantGroups\ irrelevant groups. In particular, a mixed group has both secure snippets and insecure ones, which are similar to each other. Although two filters were used (see Section~\ref{sec:extract}), 64\% of the clone groups from refined data were still irrelevant to security,
which evidences the difficulty of precisely identifying security implementation with Baker.

The clone groups cover \SecureInstances\ secure snippets, \InsecureInstances\ insecure ones, and \IrrelevantInstances\ irrelevant ones. 
When mapping these snippets to the answer posts (which contain them), we identified \SecureAnswers\ secure answers, \InsecureAnswers\ insecure ones, and \IrrelevantAnswers\ irrelevant ones. One answer can have multiple snippets of different clone groups. Therefore, 
we consider a post ``insecure'' if it contains any labeled insecure code. A post was labeled ``secure'' if it has no insecure snippet but at least one secure snippet. 
If a post does not contain any (in)secure snippet, it is labeled as ``irrelevant''. 

\subsection{Verifying the Prevalence of Sampled Posts}
\label{sec:cdf}

\begin{figure}
\vspace{-0.5em}
\centering
\includegraphics[width=8cm]{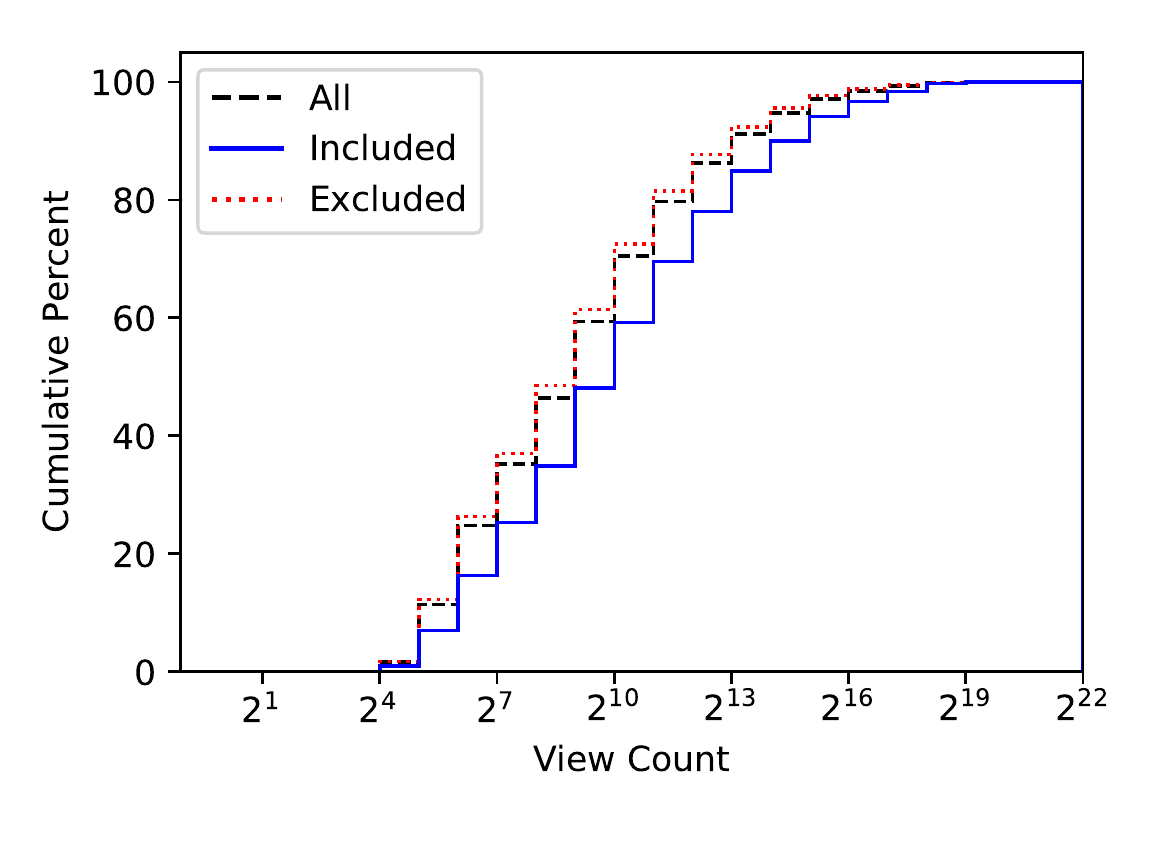}
\vspace{-1.5em}
\caption{CDFs of view count among the included answers, excluded ones, and all answers related to Baker's output}
\label{fig:cdf}
\vspace{-1em}
\end{figure}

To check whether our clone-based approach actually included representative SO posts, we separately computed the cumulative distribution functions   (CDF)~\cite{Gentle:2009} of view count for 
the included \AllAnswers\ posts (as mentioned in Table~\ref{tab:label}), the excluded \ExcludedAnswers\ posts, 
and the complete set of 23,224 posts identified by Baker.
As shown in Fig.~\ref{fig:cdf}, the ``\textit{included}'' curve is beneath the ``\textit{all}'' and ``\textit{excluded}'' curves. This shows that the highly viewed answers take up a higher percentage in our sample set than the excluded answers. 

\section{Major Findings}
\label{sec:evaluation}
In this section, we present our results and discuss the main findings regarding our stated research questions.

\subsection{Popularity of Secure and Insecure Code Suggestions}
\label{sec:rq1}

\begin{figure}
\centering
\includegraphics[width=8cm]{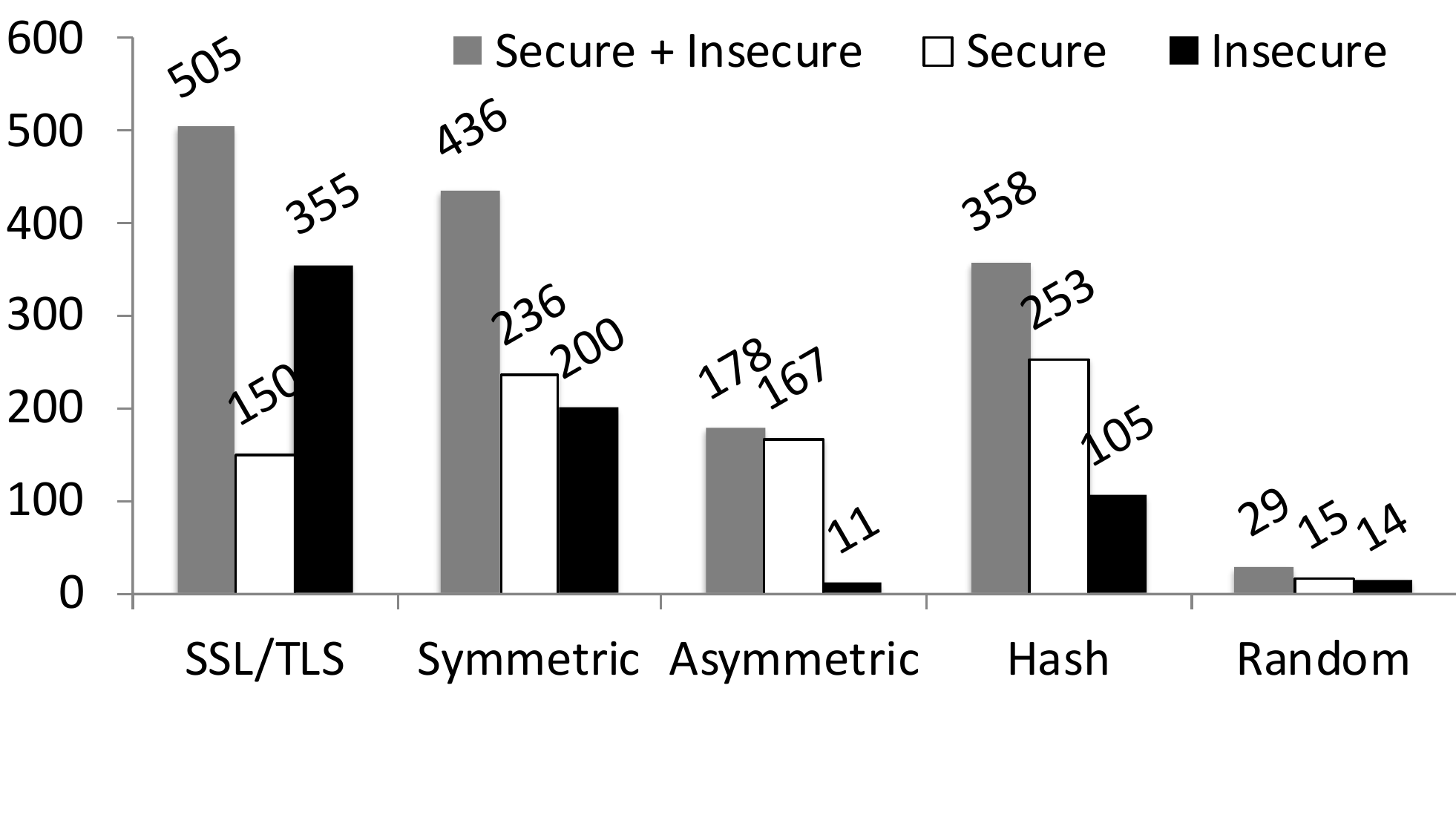}
\vspace{-2em}
\caption{The distribution of posts among different categories}
\label{fig:category}
\vspace{-1em}
\end{figure}

Figure~\ref{fig:category} presents the distribution of \RelevantAnswers\ answer posts among the 5 security categories. 
Since some posts contain multiple snippets of different categories, the total number of plotted secure and insecure posts in Figure~\ref{fig:category} is \SumCategoryAnswers, slightly larger than \RelevantAnswers. 
Among the categories, \emph{SSL/TLS} contains the most posts (34\%), 
while \emph{Random} has the fewest posts (2\%). 
Two reasons can explain such a distribution. 
First, developers frequently use or are more concerned about APIs of \emph{SSL/TLS}, \emph{Symmetric}, and \emph{Hash}. 
Second, the criteria we used to label code contain more diverse rules for the above-mentioned three categories, so we could identify more instances of such code. 

There are many more insecure snippets than secure ones in the \emph{SSL/TLS}  (\InsecureAnswersSSL\ vs.~\SecureAnswersSSL) category, indicating that 
developers should be quite cautious when searching for such code.
Meanwhile, secure answers dominate the other categories, accounting for 94\% of 
\emph{Asymmetric} posts, 71\% of \emph{Hash} posts, 54\% of \emph{Symmetric} posts, and 52\% of \emph{Random} posts. 
However, notice that across these 4 categories, only 67\% of the posts are secure;  
that is, considerable room for error remains. 

\begin{tcolorbox}
	\textbf{Finding 1:}
	\emph{\InsecureAnswers\ out of the \RelevantAnswers\ inspected answer posts (45\%) are insecure, meaning that insecure suggestions popularly exist on \SO. 
	Insecure answers dominate, in particular, the SSL/TLS category. }
\end{tcolorbox}

\begin{figure}
\includegraphics[width=8cm]{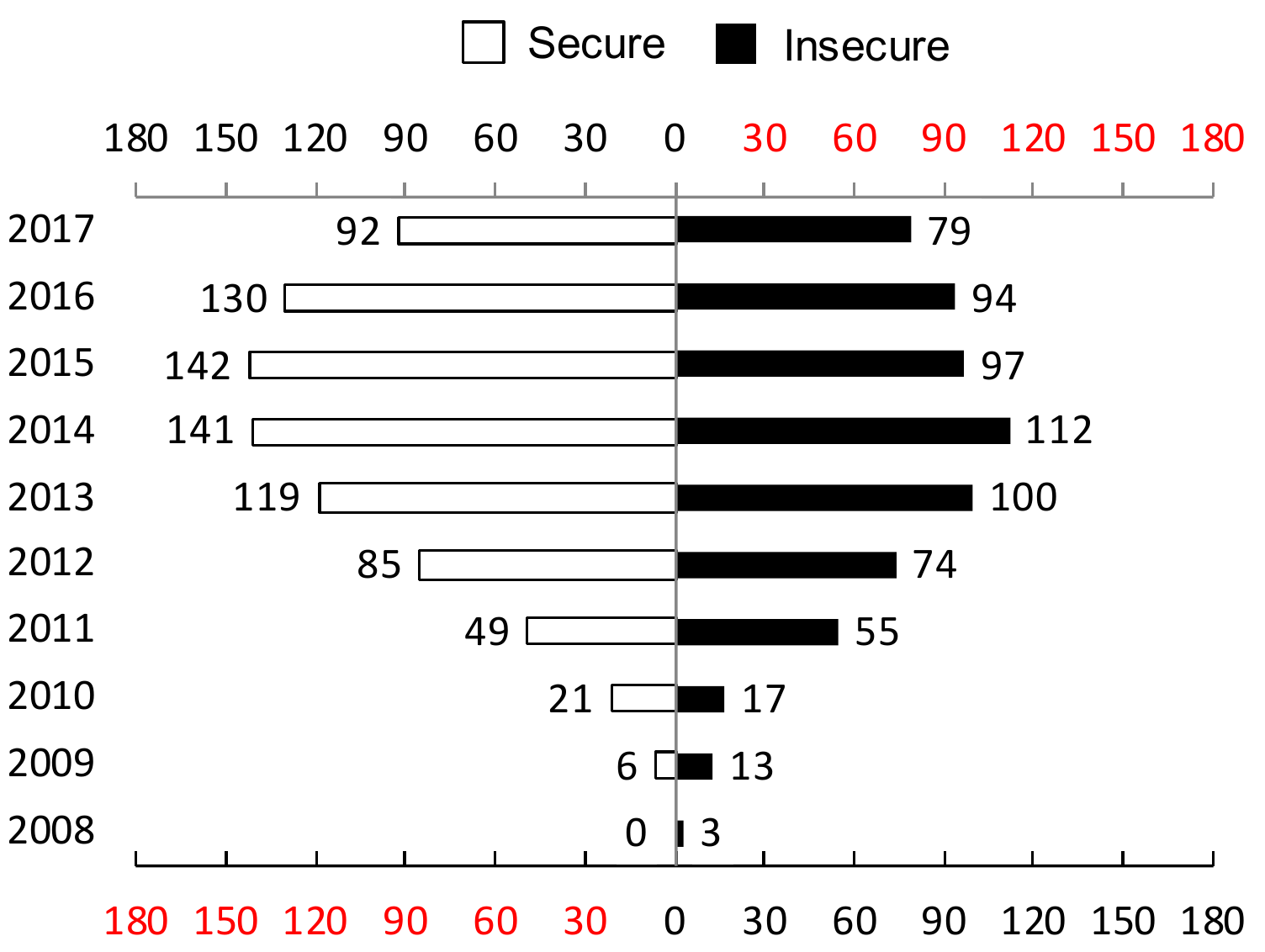}
\caption{The distribution of posts over during 2008-2017}
\label{fig:time}
\vspace{-1.5em}
\end{figure}
To explore the distribution of secure and insecure answers over time, we clustered answers based on their timestamps. 
As shown in Figure~\ref{fig:time}, both types of answers increased year-by-year from 2008 to 2014, and decreased in 2015-2017. This may be because \so reached its saturation for Java security-related discussions in 2014-2015.  
In 2008, 2009, and 2011, insecure answers were posted more often than secure ones, taking up 53\%-100\% of the sampled data of each year. 
For the other years, secure posts constitute the majority within the yearly sampled data set, accounting for 53\%-59\%. 

To further determine whether older posts are more likely to be insecure, we considered 
post IDs as logical timestamps. We applied a Mann-Whitney U test
(which does not require normally distributed data~\cite{Fay2010}), and calculated the Cliff's delta size (which measures the difference's magnitude~\cite{MACBETH2011}).  
The resulting $p$-value is 0.02, with Cliff's $\Delta$=0.07. It means that secure answers are significantly more recent than insecure ones, but the effect size is negligible. 

Two reasons can explain this finding. First, some vulnerabilities were recently revealed.  
Among the 17 insecure posts in 2008 and 2009, 6 answers use MD5, 6 answers trust all certificates, and 4 answers use TLS 1.0. 
However, these security functions were found
broken in 2011-2012~\cite{flame,Georgiev:2012,Fahl:2012,duong2011}, which made the answers obsolete and insecure.  
Second, some secure answers were posted to correct insecure suggestions. For instance, we found a question inquiring about fast and simple string encryption/decryption in Java~\cite{secure_after_insecure}. The accepted answer in 2011 suggested DES---an insecure symmetric-key algorithm. 
Later, various comments pinpointed the vulnerability and a secure answer was provided in 2013.

\begin{figure}
    \centering
    \includegraphics[width=8cm]{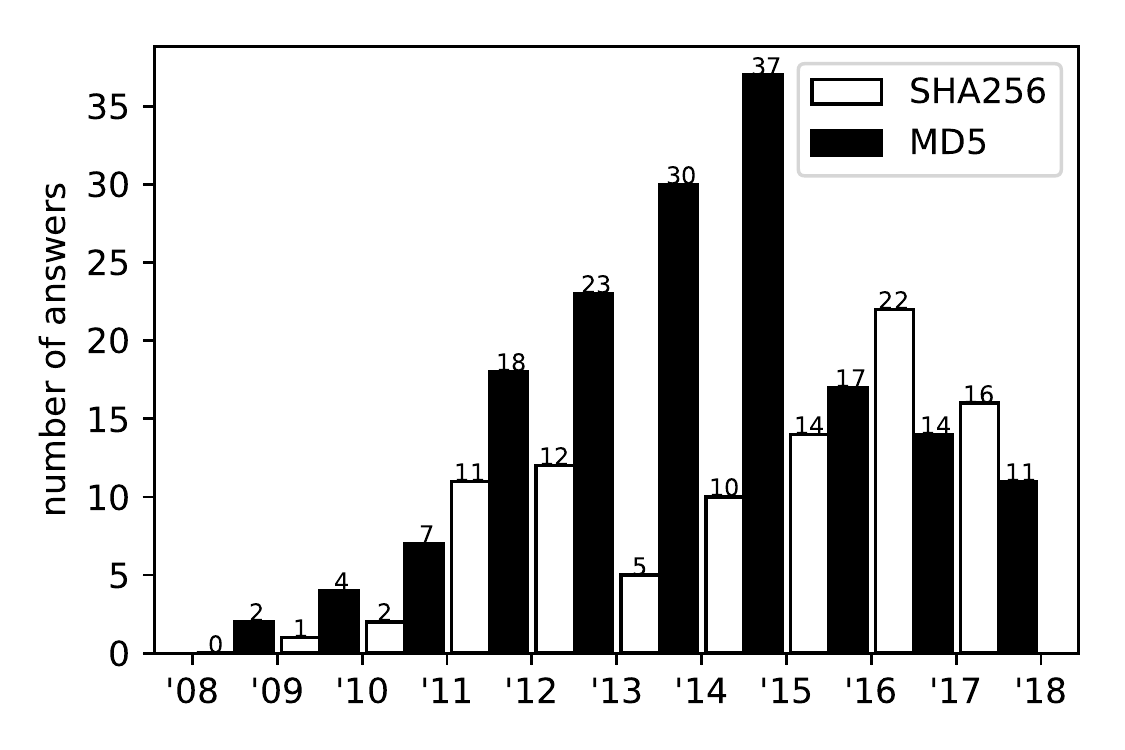}
    \vspace{-1em}
    \caption{Distributions of MD5 and SHA256 related posts}
    \label{fig:md5}
    \vspace{-1.5em}
\end{figure}

Note that there can be a significant lag until the community adopts new  secure technologies, and phases out technologies known to be insecure. 
Although MD5's vulnerability was exploited by Flame malware in 2012~\cite{flame}, as shown in Fig.~\ref{fig:md5}, 
MD5 was still popularly suggested afterwards, obtaining a peak number of related answers in 2014.

\begin{tcolorbox}
	\textbf{Finding 2:}
	\emph{Insecure posts led the sampled data in 2008-2011, while secure ones were dominant afterwards. 
	Older security-related posts are less reliable, likely because recently revealed vulnerabilities outdated older suggestions. We found only few secure answers suggested to correct outdated, insecure ones.  
	}
\end{tcolorbox} 

\begin{table*}
\caption{Comparison between secure and insecure posts}
\label{tab:all}
\vspace{-0.5em}
\scriptsize
\centering
\begin{tabular}{C{1.5cm}|C{2cm} R{2cm}R{2cm}R{2cm}R{2cm}R{2cm} }
\toprule
& & \textbf{Score} & \textbf{Comment count} & \textbf{Reputation} & \textbf{Favorite count} & \textbf{View count}\\
\toprule
\multirow{4}{*}{\textbf{All}} & \textbf{Secure mean} 
& 5 & 2 & 18,654 & 8 & 18,713\\ \cline{2-7}
& \textbf{Insecure mean} 
& 14 & 3 & 14,678 & 15 & 36,580\\ \cline{2-7}
& \textbf{p-value} 
& 0.97 & 0.02& 0.02& 0.09& 1.5e-3\\ \cline{2-7}
& \textbf{Cliff's $\Delta$}
& - & 0.07 (negligible) & 0.07 (negligible) & - & 0.10 (negligible)\\ 
\bottomrule
\multirow{2}{*}{\textbf{Category 1: }} & \textbf{Secure mean} 
& 7 & 2 & 14,447 & 9 & 21,419\\ \cline{2-7}
& \textbf{Insecure mean} 
& 18 & 3 & 15,695 & 19 & 37,445\\ \cline{2-7}
\multirow{2}{*}{\textbf{SSL/TLS}}& \textbf{p-value} 
& 0.24 & 3.3e-4 & 0.42 & 0.86 & 0.31\\ \cline{2-7}
& \textbf{Cliff's $\Delta$}
& - & 0.20 (small) & - & - & -\\ 
\bottomrule
\multirow{2}{*}{\textbf{Category 2: }} & \textbf{Secure mean} 
& 5 & 3 & 19,347 & 7 & 16,232\\ \cline{2-7}
& \textbf{Insecure mean} 
& 7 & 3 & 10,057 & 6 & 16,842\\ \cline{2-7}
\multirow{2}{*}{\textbf{Symmetric}}& \textbf{p-value} 
& 0.29 & 0.82 & 0.45 & 0.36 & 0.10\\ \cline{2-7}
& \textbf{Cliff's $\Delta$}
& - & - & - & - & -\\ 
\bottomrule
\multirow{2}{*}{\textbf{Category 3: }} & \textbf{Secure mean} 
& 5 & 2 & 17,079 & 4 & 11,987\\ \cline{2-7}
& \textbf{Insecure mean} 
& 8 & 2 & 14,151 & 3 & 9,470\\ \cline{2-7}
\multirow{2}{*}{\textbf{Asymmetric}}& \textbf{p-value} 
& 0.17 & 0.45 & 0.72 & 0.95 & 0.77\\ \cline{2-7}
& \textbf{Cliff's $\Delta$}
& - & - & - & - & -\\ 
\bottomrule
\multirow{2}{*}{\textbf{Category 4:}} & \textbf{Secure mean} 
& 5 & 2 & 20,382 & 8 & 21,254\\ \cline{2-7}
& \textbf{Insecure mean} 
& 14 & 2 & 20,018 & 22 & 74,482\\ \cline{2-7}
\multirow{2}{*}{\textbf{Hash}}& \textbf{p-value} 
& 0.26 & 0.78 & 0.18 & 0.20 & 0.07\\ \cline{2-7}
& \textbf{Cliff's $\Delta$}
& - & -  & - & - &  -\\ 
\bottomrule
\multirow{2}{*}{\textbf{Category 5:}} & \textbf{Secure mean}
 & 1 & 3 & 33,517 & 0 & 1,031\\ \cline{2-7}
& \textbf{Insecure mean} 
& 21 & 6 & 17,202 & 31 & 56,700\\ \cline{2-7}
\multirow{2}{*}{\textbf{Random}}& \textbf{p-value} 
& 0.04 & 0.02 & 0.27 & 0.02 & 0.01\\ \cline{2-7}
& \textbf{Cliff's $\Delta$}
& 0.58 (large) & 0.68 (large) & - & 0.64 (large) & 0.74 (large)\\ 
\bottomrule
\multicolumn{7}{p{16cm}}{Similar to prior work~\cite{Wang2018}, we interpreted the computed Cliff's delta value $v$ in the following way: 
(1) if $v<0.147$, the effect size is ``negligible''; 
(2) if $0.147\le v<0.33$, the effect size is ``small'; 
(3) if $0.33\le v<0.474$, the effect size is ``medium''; 
(4) otherwise, the effect size is ``large''. 
}\\
\bottomrule
\end{tabular}
\vspace{-1.5em}
\end{table*}

\subsection{Community Dynamics Towards Secure and Insecure Code}
\label{sec:rq2}

For each labeled secure or insecure post, we extracted the following information: (1) score, (2) comment count, (3) the answerer's reputation score, (4) the question's favorite count, and (5) the discussion thread's view count. 

\textbf{Comparison of Mean Values.}
Table~\ref{tab:all} compares these information categories for the \SecureAnswers\ secure posts and \InsecureAnswers\ insecure ones and applies Mann-Whitney U tests to determine significant results. 
On average, secure posts' answerers have higher  reputation (\AverReputationSecureAnswers\ vs.~\AverReputationInsecureAnswers). However, for the SSL/TLS posts, the insecure answer providers have higher reputation (15,695 vs.~14,447). 
Moreover, insecure posts have higher scores, and more comments, favorites, and views. 
Users seemed to be more attracted by insecure posts,
\emph{which is counterintuitive}. We would expect secure answers to be seen more favorable; with more votes, comments and views. 

Three reasons can explain our observation. First, software developers often face time constraints. When stuck with coding issues (\emph{e.g.}, runtime errors), developers are tempted to take insecure but simpler solutions~\cite{Meng2018:icse}. 
Take the vulnerable SSL/TLS usage in Listing~\ref{lst:ssl} for example. The insecure code was frequently suggested on \SO, and many users voted for it mainly because the code is simple and useful to resolve connection exceptions. 
Nevertheless, the simple solution essentially skips SSL verification and voids the protection mechanism. 
In comparison, a better solution should use certificates from a Certification Authority (CA) or self-signed certificates to drive the customization of \codefont{TrustManager}, and verify certificates with more complicated logic~\cite{certcreate}. 

Second, some insecure algorithms are widely supported by Java-based libraries, which can promote developers' tendency to code insecurely. For instance, up till the current version Java 9, Java platform implementations have been required to support MD5---the well-known broken hash function~\cite{md5java9}. Third, insecure posts are less recent and may have accumulated more positive scores than recent secure posts. 

\begin{tcolorbox}
	\textbf{Finding 3:}
	\emph{On average, insecure posts received higher scores, more comments, more favorites, and more views. 
It implies that (1) more user attention is attracted by insecure answers; and (2) users cannot rely on the voting system to identify secure answers. }
\end{tcolorbox} 

\textbf{Comparison of p-values and Cliff's $\Delta$.}
Table~\ref{tab:all} shows that among all posts, insecure ones obtained significantly more comments ($p=0.02$) and views ($p=1.5\mathrm{e}{-3}$), while the effect sizes are negligible. Specifically for the \emph{Random} category, insecure posts have significantly higher view counts ($p=0.01$) and the effect size is large. 
Meanwhile, 
the owners of secure answer posts have significantly higher reputation ($p=0.02$) but the magnitude is also negligible. 

\begin{figure}
\centering
\vspace{-1.1em}
\includegraphics[width=7.5cm]{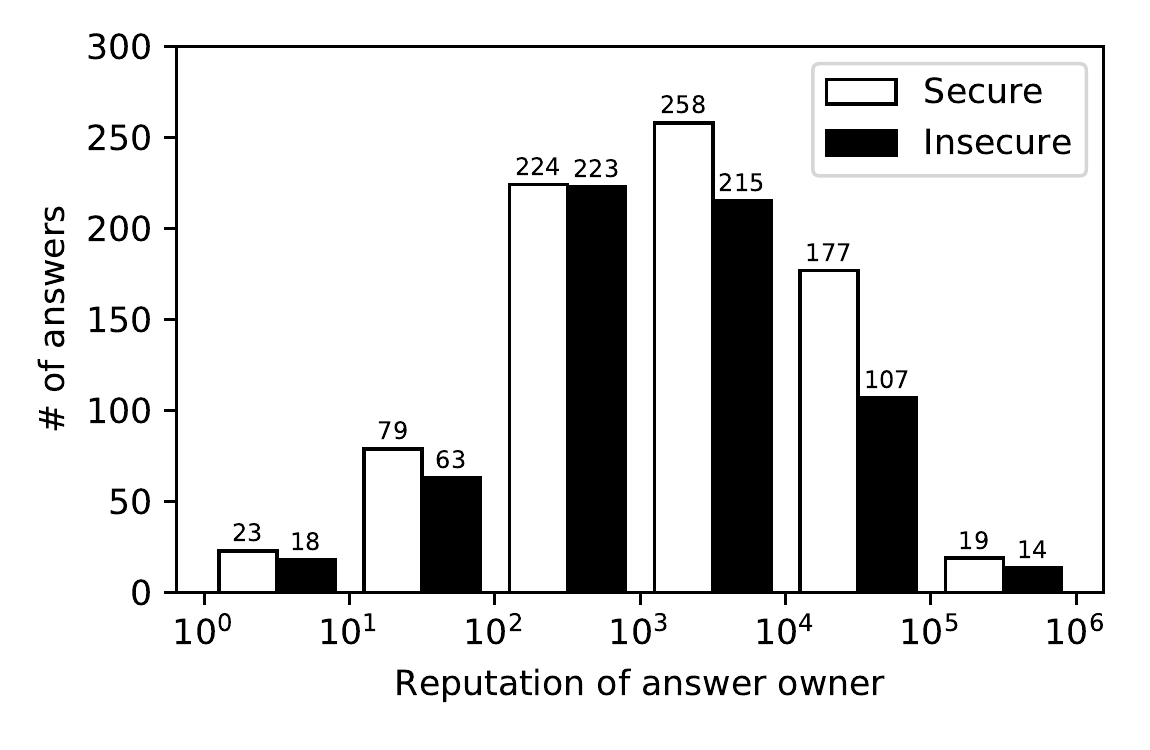}
\vspace{-1em}
\caption{The answer distribution based on owners' reputation}
\label{fig:rep}
\vspace{-1.5em}
\end{figure}

Figure~\ref{fig:rep} further clusters answers based on their owners' reputation scores. 
We used logarithmic scales for the horizontal axis, because the scores vary a lot within the range [\LowestReputation, \HighestReputation].
Overall, 
the secure and insecure answers have similar distributions among different reputation groups. For instance, most answers were provided by users with scores within [$10^2$, $10^4$), accounting for 61\% of secure posts and 68\% of insecure posts.  
Among the \RelevantAnswersByTrustedUsers\ posts by trusted users, \InsecureAnswersByTrustedUsers\ answers (34\%) are insecure and not reliable. 
One reason to explain why high reputation scores do not guarantee secure answers can be that users earned scores for being an expert in areas other than security. Responders' reputation scores do not necessarily indicate the security property of the provided answers. 
Therefore, \SO users should not blindly trust the suggestions given by highly reputable contributors.

\begin{tcolorbox}
	\textbf{Finding 4:}
	\emph{The users who provided secure answers have significantly higher reputation than the providers of insecure answers, but the difference in magnitude is negligible. Users cannot rely on the reputation mechanism to identify secure answers. }
\end{tcolorbox} 

\textbf{Comparison of accepted answers. }
It is natural for \SO users to trust accepted answers. 
Among the \RelevantAnswers\ posts, we found \AcceptedRelevantAnswers\ accepted answers (38\%). \AcceptedSecureAnswers\ accepted answers are secure, accounting for 38\% of the inspected secure posts. \AcceptedInsecureAnswers\ accepted answers are insecure, accounting for 37\% of the inspected insecure posts. 
It seems that accepted answers evenly distribute among secure and insecure posts; they are not good indicators of suggestions' security property.  

\begin{tcolorbox}
	\textbf{Finding 5:}
	\emph{Accepted answers are also not reliable for users to identify secure coding suggestions. }
\end{tcolorbox} 

\subsection{Duplication of Secure and Insecure Code}
\label{sec:rq3}

\begin{figure}
\vspace{-1.1em}
\centering
\includegraphics[width=7.5cm]{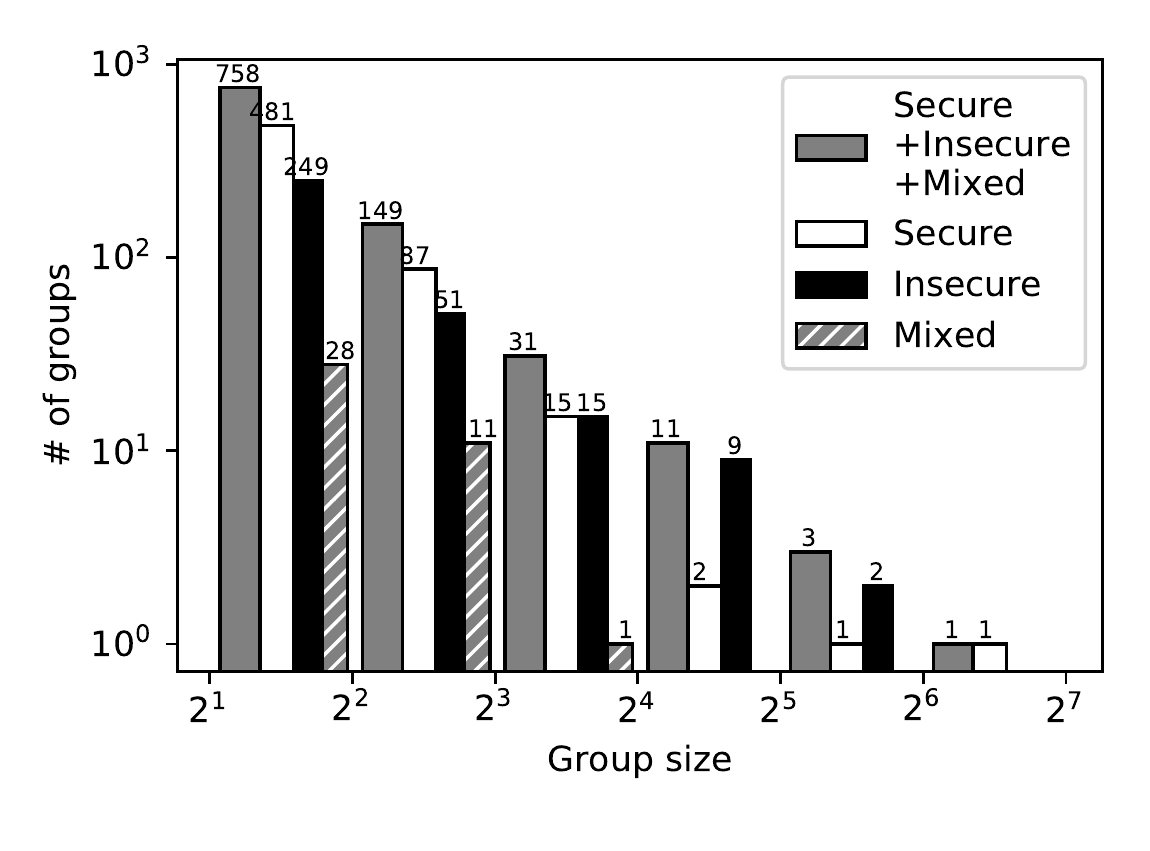}
\vspace{-1.5em}
\caption{The distribution of clone groups based on their sizes}
\label{fig:size}
\vspace{-1em}
\end{figure}

Among the \RelevantGroups\ security-related clone groups, we explored the degree of code duplication for secure clone groups, insecure groups, and mixed groups. Figure~\ref{fig:size} shows the distribution of clone groups based on their sizes.
Similar to Fig.~\ref{fig:rep}, we used logarithmic scales for both the horizontal and vertical axes. 
In Fig.~\ref{fig:size}, most clone groups are small, with 2-3 similar snippets. The number of groups decreases 
dramatically as the group size increases. 
Interestingly, within $[2^4, 2^6)$, there are more insecure groups than secure ones.
Table~\ref{tab:size} compares the sizes of secure and insecure groups. 
Surprisingly, insecure groups have significantly larger sizes than secure ones, although the difference is negligible. 
Our observations imply that \emph{the frequently mentioned code snippets on \SO are not necessarily more secure than less frequent ones}. Users cannot trust code's repetitiveness to decide the security property. 

\begin{table}
\caption{Comparison between secure and insecure groups in terms of their group sizes}
\label{tab:size}
\vspace{-0.5em}
\scriptsize
\begin{tabular}{c| R{1.7cm} R{1.7cm} R{1.2cm} R{1.7cm}}
\toprule
& \textbf{Secure groups' mean} & \textbf{Insecure groups' mean} & \textbf{p-value} & \textbf{Cliff's $\Delta$}\\ 
\toprule
\textbf{Size} & \SecureGroupMeanSize & \InsecureGroupMeanSize & \GroupSizePValue & \GroupSizeCliff\ (negligible)\\
\bottomrule
\end{tabular}
\vspace{-1.5em}
\end{table}

\begin{tcolorbox}
	\textbf{Finding 6: }
	\emph{
	Repetitiveness does not guarantee security, 
	so users cannot assume 
	a snippet to be secure simply because it is recommended many times. 
	}
\end{tcolorbox} 

To understand why there are mixed groups that contain similar secure and insecure implementations, we conducted a case study on  10 randomly selected mixed groups. Among \textit{all} these groups, secure snippets differ from insecure ones by using distinct parameters when calling security APIs.
This implies a great opportunity to build security bug detection tools that check for the parameter values of specific APIs. 

\begin{lstlisting}[caption=A clone group with both secure and insecure code, label=lst:clones]
(*\bfseries // An insecure snippet using AES/ECB to create a cipher~\cite{clone1}*)
Cipher cipher = Cipher.getInstance("AES/ECB/PKCS5Padding", "SunJCE");
Key skeySpec=KeyGenerator.getInstance("AES").generateKey();
cipher.init(Cipher.ENCRYPT_MODE, skeySpec);
System.out.println(Arrays.toString(cipher.doFinal(new byte[] { 0, 1, 2, 3 })));
(*\bfseries //A secure snippet using AES/CFB to create a cipher~\cite{clone2}*)
final Cipher cipher=Cipher.getInstance("AES/CFB/NoPadding", "SunJCE");
final SecretKey skeySpec=KeyGenerator.getInstance("AES").generateKey();
cipher.init(Cipher.ENCRYPT_MODE, skeySpec);
System.out.println(Arrays.toString(cipher.doFinal(new byte[] { 0, 1, 2, 3 })));
\end{lstlisting}

Listing~\ref{lst:clones} shows a mixed clone group, where the insecure code uses ``\codefont{AES/ECB}'' to create a cipher, and the secure code uses ``\codefont{AES/CFB}''. 
Actually, both snippets were provided by the same user, which explains why they are so similar. These answers are different because the askers inquired for different modes (ECB vs.~CFB). Although the answerer is an expert in using both APIs and has a high reputation score 27.7K, he/she did not mention anything about the vulnerability of ECB. This may imply a lack of security expertise or vulnerability awareness of highly reputable \SO users, and a well-motivated need for automatic tools to detect and fix insecure code.  

\begin{tcolorbox}
	\textbf{Finding 7: }
	\emph{
	Secure and insecure code in the same mixed group often differs 
	by passing distinct parameters to the same security APIs highlighting opportunities for automatic tools to handle security weaknesses. 
	}
\end{tcolorbox}

\subsection{Creation of Duplicated Secure and Insecure Code}
\label{sec:rq4}

We conducted two case studies to explore
why duplicated code was suggested.  


\textbf{Case Study I: Duplicated answers by different users.} 
We examined the largest secure group and largest insecure group. 
The secure group has 65 clone instances, which are similar to the code in Listing~\ref{lst:facebook}. 
These snippets were offered to answer questions on how to enable an Android app to log into Facebook. 
The questions are similar but different in terms of the askers' software environments (\emph{e.g.}, libraries and tools used) and potential solutions they tried. Among the 65 answers, only 18 (28\%) were marked as accepted answers. 
The majority of duplicated suggestions are relevant to the questions, but cannot solve the issues. 
\SO users seemed to repetitively provide ``generally best practices'',  
probably because they wanted to earn points by answering more questions. 

\begin{lstlisting}[caption=An exemplar snippet to generate a key hash for Facebook login~\cite{facebook}, label=lst:facebook]
PackageInfo info = getPackageManager().getPackageInfo("com.facebook.samples.hellofacebook", PackageManager.GET_SIGNATURES);
for (Signature signature : info.signatures) {
  MessageDigest md = MessageDigest.getInstance("SHA");
  md.update(signature.toByteArray());
  Log.d("KeyHash:", Base64.encodeToString(md.digest(), Base64.DEFAULT)); }
\end{lstlisting}

The largest insecure group contains 32 clone instances, which are similar to the code 
in Listing~\ref{lst:ssl}. The questions are all about how to implement SSL/TLS or resolve SSL connection exceptions. 13 of these answers (41\%) were accepted. We noticed that only one answer warns ``\emph{Do not implement this in production code \ldots}''~\cite{sslwarn}.
Six answers have at least one comment talking about the vulnerability.  
The remaining 25 answers include nothing to indicate the security issue.  

\textbf{Case Study II. Duplicated answers by the same users.} 
In total, \AnswerReusageUsers\ users reused code snippets to answer multiple questions. 
Among the \RelevantGroupsOfReusedAnswers\ clone groups these users produced, there are \SecureGroupsOfReusedAnswers\ secure groups, \InsecureGroupsOfReusedAnswers\ insecure groups, and 
\MixedGroupsOfReusedAnswers\ mixed groups. 
\SecureReusageUsers\ users repetitively posted secure answers, and \InsecureReusageUsers\ users posted duplicated insecure answers. Six among these users posted both secure and insecure answers. 
Most users (\emph{i.e.}, 92) only copied code once and produced two duplicates. One user posted nine insecure snippets, with seven snippets using an insecure version of TLS, and two snippets trusting all certificates. 
This user has 17.7K reputation (top 2\% overall) and is an expert in Android. By examining the user's profile, we did not find any evidence to show that the user intentionally misled people. It seems that the user was not aware of the vulnerability when posting these snippets.

To understand whether duplicated code helps answer questions, we randomly sampled 103 (of the 208) clone groups resulting in \SecureSamplePairs\ secure clone pairs, \InsecureSamplePairs\ insecure pairs, and \MixedSamplePairs\ mixed pairs. 
Unexpectedly, we found that \ReusedPairsForDiffQ\ pairs (45\%) did not directly answer the questions. 
For instance, a user posted code without reading the question and received down-vote (i.e., $-1$)~\cite{noread}.
In the other 57 cases, duplicated code was provided to answer similar or identical questions.

\begin{tcolorbox}
	\textbf{Finding 8: }
	\emph{Duplicated answers were created because (1) users asked  similar or related questions; and (2) some users blindly copied and pasted code to answer more questions and earn points. 
	However, we did not identify any user that intentionally misled people by posting insecure answers. 
	}
\end{tcolorbox}

\section{Related Work}
\label{sec:related}


\subsection{Security API Misuses}

Prior studies showed that API misuses caused security vulnerabilities~\cite{LongSoftwareVulnerabilities2005,
Fahl:2012,Georgiev:2012,Egele:2013,Lazar:2014,veracode,Yang2016,fischer:2017}. 
For instance, 
Lazar et al.~analyzed 369 published cryptographic vulnerabilities in the CVE database, and found that 83\% of them were caused by API misuses
~\cite{Lazar:2014}. 
Egele \emph{et al.}~built a static checker for six well-defined Android cryptographic API usage rules (e.g., ``Do not use ECB mode for encryption''). 
They analyzed 11,748 Android applications for any rule violation~\cite{Egele:2013}, and found 88\% of the applications violating at least one checked rule.
Instead of checking for insecure code in CVE or software products, we focused on \SO. 
Because the insecure coding suggestions on \SO can be read and reused by many  developers, they have a  profound impact on software quality.  

The research by Fischer \emph{et al.}~\cite{fischer:2017} is closely related to our work.
In their work, secure and insecure snippets from SO were used to search for code clones in Android apps. Our research is different in three aspects. First, it 
investigates the evolution and distribution of secure and insecure coding suggestions within the SO ecosystem itself.
Second, while [5] compares average score and view counts for secure and insecure snippets, they merely do this for snippets whose exact copies have migrated into apps but not for our much broader set of snippets on SO. Therefore, the dataset of [5] is not representative to evaluate the impact of severity, community's awareness, and popularity of unreliable SO suggestions on secure coding. 
Third, we conducted not only statistical testing on a comprehensive dataset to quantitatively contrast score, view count, comment count, reputation, and favorite count, but also case studies to qualitatively analyze the differences. We further explored the missing link between gamification and security advice quality on crowdsourcing platforms. 


\subsection{Developer Studies}
Researchers conducted interviews or surveys to understand  developers' security coding practices~\cite{xie_why_2011,balebako_improving_2014,Nadi:2016,Acar:2016}. 
For example, Nadi \emph{et al.} surveyed 48 developers and revealed that developers found it difficult to use cryptographic algorithms correctly~\cite{Nadi:2016}.
Xie \emph{et al.} interviewed 15 developers, and found that (1) most developers had reasonable knowledge about software security, but (2) they did not consider security assurance as their own  responsibility~\cite{xie_why_2011}.  
Acar \emph{et al.} surveyed 295 developers and conducted a lab user study with 54 students and professional Android developers~\cite{Acar:2016}. 
They observed that most developers used search engines and \SO to address security issues. 
These studies inspired us to explore how much we can trust the crowdsourced knowledge of security coding on \SO. 

\subsection{Empirical Studies on Stack Overflow}

Researchers conducted various studies on \SO~\cite{mamykina_design_2011,Bosu2013,Barua2014,Yang2016,rahman,Meng2018:icse,zhang_are_2018}. Specifically,   
Zhang \emph{et al.}~studied the JDK API usage recommended by \SO, and observed that 31\% of the studied posts misused APIs~\cite{Meng2018:icse}. 
Meng \emph{et al.} manually inspected 503 \SO discussion threads related to Java security~\cite{Meng2018:icse}. They revealed various secure coding challenges (\emph{e.g.}, hard-to-configure third-party frameworks) and vulnerable coding suggestions (\emph{e.g.}, SSL/TLS misuses). 
Mamykina \emph{et al.}~revealed several reasons (\emph{e.g.}, high response rate) to explain why \so is one of the most visible venues for expert knowledge sharing~\cite{mamykina_design_2011}.
Vasilescu \emph{et al.}~studied the associations between \SO and GitHub, and found that GitHub committers usually ask fewer questions and provide more answers~\cite{Vasilescu2013}.
Bosu \emph{et al.}~analyzed the dynamics of reputation building on SO, and found that answering as many questions as possible can help users quickly earn reputation~\cite{Bosu2013}. 



In comparison, our paper quantitatively and qualitatively analyzed secure and insecure \SO suggestions in terms of (1) their popularity, (2) answerers' reputations, (3) the community's feedback to answers  (\emph{e.g.}, votes and comments), and (4) the degree and causes of duplicated answers.  
We are not aware of any prior work that analyzes \SO posts in these aspects.  


\subsection{Duplication Detection Related to SO or Vulnerabilities}

Researchers used clone detection to identify duplication within \so or between \so and software products
~\cite{Chen:2015,ahasanuzzaman_mining_2016,An2017,zhang_detecting_2017,yang_stack_2017}. 
Specifically, Ahasanuzzaman \emph{et al.}~detected duplicated \SO questions with machine learning~\cite{ahasanuzzaman_mining_2016}. 
An \emph{et al.} compared code between \SO and Android apps and observed unethical code reuse phenomena on \SO~\cite{An2017}. 
Other researchers used static analysis to detect vulnerabilities caused by code cloning~\cite{Pham:2010:DRS,Jang:2012:ReDeBug,Li:2016:VAV,Kim2017:VUDDY}. For instance, Kim et al.~generate a fingerprint for each Java method to efficiently search for clones of a given vulnerable snippet~\cite{Kim2017:VUDDY}. 
Different from prior work, we did not invent new clone detection techniques or compare code between \so and software projects. 
We used clone detection to (1) sample crawled security-related code, and (2) explore why \SO users 
posted similar code to answer questions.

\section{Our Recommendations}
\label{sec:suggestion}

By analyzing \SO answer posts relevant to Java-based security library usage, we observed the wide-spread existence of insecure code. 
It is worrisome to learn that \SO users cannot rely on either the  reputation mechanism or voting system to infer
an answer's security property, 
A recent Meta Exchange discussion thread also shows the frustration of \SO developers to keep outdated security answers up to date~\cite{metaexchange}. Below are our recommendations based on this analysis. 

\paragraph{\textbf{For Tool Builders}} Explore approaches that accurately and flexibly detect and fix security bugs. 
Although a few tools identify security API misuses through static program analysis or machine learning~\cite{Egele:2013,He2015,Rahaman:2017,fischer:2017}, they are unsatisfactory due to the (1) hard-to-extend API misuse patterns hardcoded in tools, and (2) hard-to-explain machine learning results.
People report vulnerabilities and patches on CVE and in security papers. Tools like LASE~\cite{Meng2013:Lase} were built to (i) generalize program transformation from concrete code changes, and (ii) leverage the transformation to locate code for similar edits. If tool developers can extend such tools to compare secure-insecure counterparts, they can automatically fix vulnerabilities in a flexible way. 


\paragraph{\textbf{For \SO Developers}} Integrate static checkers to scan existing corpus and \SO posts under submission. Automatically add warning messages or special tags to any post that has vulnerable code. Encourage moderators or trusted users to exploit clone detection technologies in order to efficiently detect and remove both duplicated questions and answers. Such deduplication practices will not only save users' time and effort of reading/answering useless duplicates, but also mitigate the misleading consensus among multiple similar insecure suggestions. 
As user profiles include top tags to reflect the frequently asked/answered questions by users. instead of accumulating a single reputation score for each user, SO developers can compute one score for each top tag to better characterize users' expertise. 


\paragraph{\textbf{For Designers of Crowdsourcing Platforms}}
Provide incentives to users for downvoting or detailing vulnerabilities and suggesting secure alternatives. 
Introduce certain mechanisms to encourage owners of outdated or insecure answers to proactively archive or close such posts. 
We expect researchers from the usable security and HCI domain to evaluate and test new design patterns that integrate security evaluation in the gamification approach.



\section{Threats to Validity}
\label{sec:threats}

\paragraph{Threat to External Validity}
This study labels insecure code based on the Java security rules summarized by prior work~\cite{fischer:2017}, so our studied insecure snippets are limited to Java code and these rules. Since we used the state-of-the-art insecurity criteria, our analysis revealed as diverse insecure code as possible. In the future, we plan to identify more kinds of insecure code by considering different programming languages and exploiting multiple vulnerability detection tools~\cite{flawFinder,checkmarx}.


\paragraph{Threat to Construct Validity}
Although we tried our best to accurately label code, our analysis may be still subject to human bias and cannot scale to handle all crawled data or more security categories. We conservatively assume a snippet to be secure if it does not match any given rule. However, it is possible that some labeled secure snippets actually match the insecurity criteria not covered by this study, or will turn out to be insecure when future attack technologies are created. 
We concluded that insecure answers are popular on \SO and gain high scores, votes, and views. 
Even if the labels of some existing secure answers will be corrected as insecure in the future, our conclusion generally holds. 

\paragraph{Threat to Internal Validity}
We leveraged clone detection to sample the extracted code snippets and reduce our manual analysis workload. Based on code's occurrence repetition, clone detection can ensure the representativeness of sampled data. However, the measurement on 
a sample data set may be still different from that of the whole data set. Once we build automatic approaches to precisely identify security API misuses, we can resolve this threat.

\section{Conclusion}
\label{sec:conclusion}

We aimed to assess the reliability of the crowdsourced knowledge on security implementation. Our analysis of 1,429 answer posts on \SO revealed 3 insights. 
\begin{enumerate}
\item In general secure and insecure advices more or less balance each other (55\% secure and 45\% insecure). Users without security knowledge may heavily rely on the community to provide helpful feedback in order to identify secure advice. Unfortunately, we found the community's feedback to be almost useless. 
For certain cryptographic API usage scenarios, the situation is even worse: insecure coding suggestions about SSL/TLS dominate the available options. This is particularly alarming as SSL/TLS is one of the most common use cases in production systems according to prior work~\cite{fischer:2017}.

\item The reputation mechanism and voting system popularly used in crowdsourcing platforms turn out to be powerless to remove or discourage insecure suggestions. Insecure answers were suggested by people with high reputation and widely accepted as easy fixes for programming errors. On average, insecure answers received more votes, comments, favorites, and views than secure answers.
As a countermeasure, security evaluation can be included in the voting and reputation system to establish missing incentives for providing secure and correcting insecure content.

\item 
When users are motivated to earn reputation by answering more questions, the platform encourages contributors to provide duplicated, less useful, or insecure coding suggestions. Therefore, with respect to security, \so{}'s gamification approach counteracts its original purpose as it promotes distribution of secure and insecure code. Although we did not identify any malicious user that misuses \SO to propagate insecure code, we do not see any mechanism designed to prevent such malicious behaviors, either.

\end{enumerate}

When developers refer to crowdsourced knowledge as one of the most important information resources, 
it is crucially important to enhance the quality control of crowdsourcing platforms. This calls for a strong collaboration between developers, security experts, tool builders, educators, and  platform providers. By educating developers to contribute high-quality security-related information, and integrating vulnerability and duplication detection tools into platforms, we can improve software quality via crowdsourcing. Our future work is focused on building the needed tool support. 

\section*{Acknowledgment}
We thank reviewers for their insightful comments. We also thank Dr.~Kurt Luther for his valuable feedback. 
\bibliographystyle{IEEEtran}
\bibliography{main_clone_so} 

\end{document}